\documentclass[prl,twocolumn,showpacs,superscriptaddress,amsmath,amssymb]{revtex4}

\usepackage{graphicx}
\usepackage{dcolumn}
\usepackage{bm}

\begin{document}

\title{Spin Gap and Resonance at the Nesting Wavevector in Superconducting FeSe$_{0.4}$Te$_{0.6}$}

\author{Yiming Qiu}
\affiliation{NIST Center for Neutron Research, National Institute of Standards
and Technology, Gaithersburg, MD 20899, USA} 
\affiliation{Department of Materials Science and Engineering, University of Maryland, College Park, MD 20742, USA}
\author{Wei Bao}
\email{wbao@ruc.edu.cn}
\affiliation{Department of Physics, Renmin University of China, Beijing 100872, China}
\author{Yang Zhao}
\affiliation{Department of Physics and Astronomy, Johns Hopkins University, Baltimore, MD 21218， USA}
\author{Collin Broholm}
\affiliation{Department of Physics and Astronomy, Johns Hopkins University, Baltimore, MD 21218， USA}
\affiliation{NIST Center for Neutron Research, National Institute of Standards
and Technology, Gaithersburg, MD 20899, USA} 
\author{V. Stanev}
\author{Z. Tesanovic}
\affiliation{Department of Physics and Astronomy, Johns Hopkins University, Baltimore, MD 21218， USA}
\author{Y.C. Gasparovic}
\affiliation{NIST Center for Neutron Research, National Institute of Standards
and Technology, Gaithersburg, MD 20899, USA}
\affiliation{Department of Materials Science and Engineering, University of Maryland, College Park, MD 20742, USA}
\author{S. Chang}
\affiliation{NIST Center for Neutron Research, National Institute of Standards and Technology, Gaithersburg, MD 20899, USA}
\author{Jin Hu}
\author{Bin Qian}
\affiliation{Department of Physics, Tulane University, New Orleans, LA 70118， USA}
\author{Minghu Fang}
\affiliation{Department of Physics, Tulane University, New Orleans, LA 70118， USA}
\affiliation{Department of Physics, Zhejian University, Hangzhou 310027, China}
\author{Zhiqiang Mao}
\affiliation{Department of Physics, Tulane University, New Orleans, LA 70118， USA}

\date{\today}

\begin{abstract}
Neutron scattering is used to probe magnetic excitations in $\rm FeSe_{0.4}Te_{0.6}$ ($T_c$=14 K). Low energy spin fluctuations are found with a characteristic wave vector $(\frac{1}{2}\frac{1}{2}L)$ that corresponds to Fermi surface nesting and differs from ${\bf Q}_m=(\delta 0 \frac{1}{2})$ for magnetic ordering in $\rm Fe_{1+y}Te$. A spin resonance with $\hbar\Omega_0=6.5$~meV~$\approx 5.3 k_BT_c$ and $\hbar\Gamma=1.25$~meV develops in the superconducting state from a normal state continuum. We show that the resonance is consistent with a bound state associated with $s_\pm$ superconductivity and imperfect quasi-2D Fermi surface nesting.
\end{abstract}


\maketitle
\begin{figure}[b]
\includegraphics[width=86mm,angle=0]{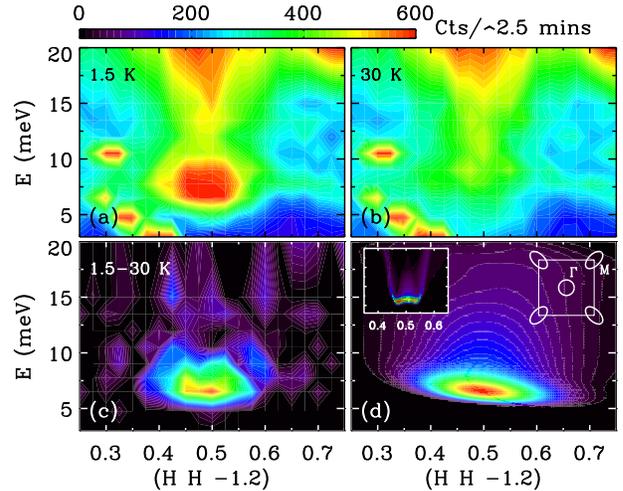}
\caption{(color) Spin excitation spectrum as a function of ${\bf Q}=(H,H,-1.2)$ and energy at (a) 1.5 K and (b) 30 K. (c) The difference between the 1.5 K and 30 K spectra. The intensity in (c) is multiplied by a factor of 2 so that the same intensity scale at the top is used for (a)-(c). (d) Resolution convolved theoreticaly difference intensity from a simplified two-band model extracted from ARPES measurements \cite{B011299}. The insets show the resolution free theoretical intensity difference (left) and the normal state Fermi surface employed (right). 
}
\label{fig1}
\end{figure}

The recent discovery of superconductivity in oxypnictides of the form $R$FeAsO (1111) \cite{Kamihara2008} has triggered a burst of
scientific activity. Subsequently discovered superconductivity in suitably doped BaFe$_2$As$_2$ (122) \cite{A054630}, LiFeAs \cite{A064688}, and
FeSe (11) \cite{A072369} indicates a crucial role for the
shared FeAs or FeSe antifluorite layer. The theoretical electronic structure is indeed dominated at the Fermi level by contributions from this layer \cite{B012149} and density functional theory \cite{A033286} successfully predicts, as a consequence of Fermi surface nesting, the observed antiferromagnetic order in the  1111 \cite{A040795} and 122 type parent compounds \cite{A062776,A071525}. In contrast the magnetic parent compounds of 11-type superconductors order with a tunable antiferromagnetic vector ${\bf Q}_m=(\delta 0\frac{1}{2})$ \cite{A092058}, with an in-plane component that is rotated by 45$^o$ with respect to the ${\bf Q}_n=(\frac{1}{2}\frac{1}{2})$ nesting vector connecting the $\Gamma$ and ${\rm M}$ points [see inset to Fig.~\ref{fig1}(d)].

Recently, a spin resonance was discovered at the antiferromagnetic nesting vector in BaFe$_2$As$_2$-derived superconductors \cite{A073932,A114755,A121354}. This raises several important questions: Does a spin resonance generally exist for iron pnictide superconductor and for Fe(Se,Te) in particular, where in reciprocal space is it to be found? In this letter we report that superconducting Fe(Se,Te) {\em does} exhibit a spin resonance though not at ${\bf Q}_m$ but at the $\Gamma-M$ Fermi surface nesting vector, ${\bf Q}_n$. This indicates a common form of superconductivity in the 122 and 11 families of iron based superconductors and brings into view a striking unifying feature of a wide range of unconventional superconductors proximate to magnetism: They exhibit a spin resonance at an energy $\hbar\Omega_0$ that scales with $k_BT_c$ \cite{UemuraII} and a commensurate wave vector that reverses the sign of the superconducting order parameter. 

Single crystals of FeSe$_{0.4}$Te$_{0.6}$ were grown by a flux method. Growth methods and bulk properties are reported elsewhere \cite{B040824}. Bulk superconductivity in the sample labeled SC1 appears through a sharp transition in resistivity, magnetic susceptibility and heat capacity with an onset at $T_c\approx 14$ K. The lattice parameters of the tetragonal $P4/nmm$ unit cell are $a=b=3.802$\~\AA\ and
$c=6.061$~\AA\ at room temperature. Five crystals, weighing $\sim$2 g each, were mutually aligned to increase counting efficiency. Magnetic neutron scattering measurements were performed using the thermal (BT7) and cold (SPINS) neutron triple-axis spectrometers at NIST. The sample temperature was controlled by a pumped $^4$He cryostat. As opposed to experiments on samples containing 8\% excess Fe  \cite{A092058}, no low energy magnetic signals were detected at the antiferromagnetic wavevector ${\bf Q}_m=(\delta 0\frac{1}{2})$ in FeSe$_{0.4}$Te$_{0.6}$. Therefore, we will 
concentrate on results from scans in the ($HHL$) reciprocal plane from BT7, using the fixed $E_f$=14.7 meV configuration. Measurements with better resolution and in a 7 T cryomagnet were conducted at SPINS using $E_f$=4.2 meV. Pyrolytic Graphite (PG) and cooled Be was used to reject order contamination on BT7 and SPINS respectively and both instruments employ PG to monochromate the incident beam and analyze the scattered beam.

Figure~\ref{fig1}(a) and (b) show the spin excitation spectrum of FeSe$_{0.4}$Te$_{0.6}$,
combining 10 different constant energy scans through the in-plane nesting vector $(\frac{1}{2}\frac{1}{2})$, at temperature $T=1.5$~K and 30 K, respectively. The temperature independent, sharp spurions in (a) and (b) cancel in the difference spectrum (Fig. 1(c)). An intense ``resonance'' sharply defined both in energy and in-plane momentum appears below $T_c$, above the normal state ridge-like continuum at the nesting vector ${\bf Q}_n$ rather than at the wave vector ${\bf Q}_m$ of the antiferromagnetic parent compound. Correspondingly we note that while the ``parent''
non-superconducting heavy fermion compound CeRhIn$_5$ orders in an incommensurate antiferromagnetic structure \cite{bao00a},
the resonance in superconducting CeCoIn$_5$ appears at the commensurate wavevector associated with $d_{x^2-y^2}$ superconductivity \cite{Co115_stock}.

\begin{figure}
\includegraphics[width=82mm,angle=0]{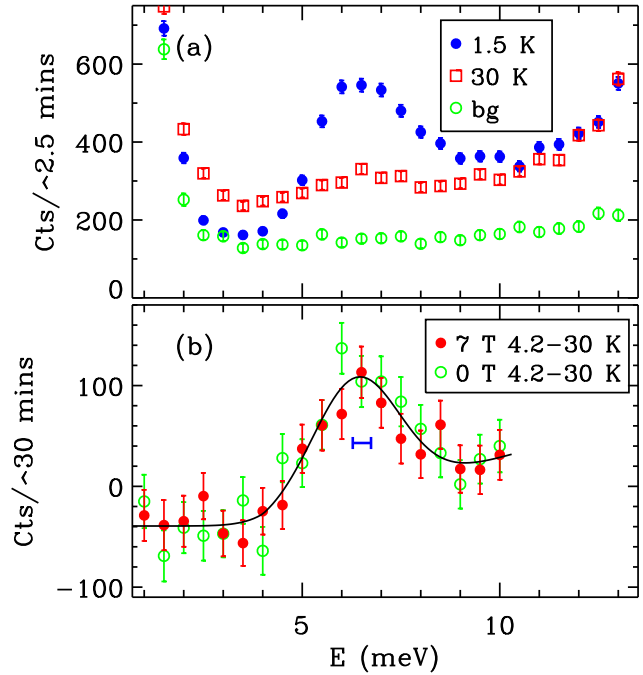}
\caption{(color online) (a) Constant ${\bf Q}=(0.46,0.46,0.65)$ scan at 1.5 K and 30 K, measured at BT7. The sample-turned background measured at 30 K is shown by green open circles.
(b) The difference intensity of the const-${\bf Q}$=$(\frac{1}{2}\frac{1}{2}L)$ scans measured at 4.2 and 30 K using SPINS, with and without an applied 7 T magnetic field. The solid line is a guide to the eye. The blue bar indicates the instrumental energy resolution (FWHM). }
\label{fig2}
\end{figure}
Figure~\ref{fig2}(a) shows constant-${\bf Q}$ scans through the resonance above and below $T_c$, together with
measured background. The spectrum appears to be gapless in the normal state as measured at 30 K with a weak "knee" at the resonance energy. 
The normal state data in Fig.~\ref{fig1}(b)
is similar to data from paramagnetic and metallic V$_2$O$_3$ (see Fig.~2(b) of Ref.~[\onlinecite{bao96c}]) indicating spin fluctuations resulting from Fermi surface nesting.
At 1.5 K, a full spin gap is opened at low energies as spectral weight concentrates
in a ``resonance'' peak at $\hbar\Omega_0=6.51(4)$ meV.
Higher resolution constant-${\bf Q}$ scans measured using cold neutrons are shown in 
Fig.~\ref{fig2}(b). Here the SPINS spectrometer was arranged so the line of intensity from ${\bf Q}=(0.5,0.5,-0.796)$ to $(0.5,0.5,-1.804)$  was collected by the focusing analyzer during the scan. The resonance peak is much wider than the full width at half maximum (FWHM) instrumental resolution, 0.48 meV. The resolution-corrected half width $\hbar\Gamma= 1.25(5)$ meV may indicate a finite lifetime of the resonant spin fluctuations, imperfect nesting, or broadening due to disorder on the Se/Te site.

\begin{figure}
\includegraphics[width=86mm,angle=0]{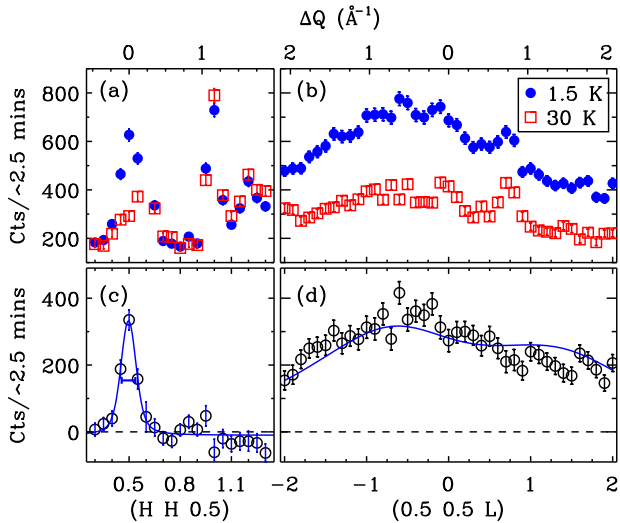}
\caption{(color online)  Constant $\hbar\omega=6.5$ meV scans (a) along the ($HH\frac{1}{2}$) direction, and (b) along the ($\frac{1}{2}\frac{1}{2}L$) direction, showing the quasi-two-dimensionality of the spin resonance. (c)-(d) The difference between the 1.5 K and 30 K scans. In (c) the solid line is a fit to a resolution convolved lorentzian, and the horizontal bar represents the FWHM of the resolution. In (d) the line is product of the Fe magnetic form factor squared and the projection of the instrumental resolution volume along $\bf c$.}
\label{fig3}
\end{figure}
To determine the spatial correlations associated with the resonance, constant energy scans
were performed in its vicinity. Figure~\ref{fig3}(a) shows a basal plane scan at the resonance energy
covering a full Brillouin zone. Weak intensity at ${\bf Q}=(\frac{1}{2}\frac{1}{2}\frac{1}{2})$ and $T=30$ K is strongly enhanced in the superconducting state at $T=1.5$~K. 
The net enhancement is shown by the difference data in Fig.~\ref{fig3}(c). Spurions exists at both temperatures but these cancel in the difference plot. The horizontal bar indicates the FWHM instrumental resolution. Based on the calculated resolution function, the deconvolved half width at half maximum is $0.023(5)\times \sqrt{2}\times a^*$, indicating a correlation length of 19(4) $\AA$ or 7(1) Fe-Fe lattice spacings. Figure~\ref{fig3}(b)
shows a scan in the inter-plane direction above and below $T_c$, with the difference in (d).
As for quasi-two-dimensional BaFe$_{1.84}$Co$_{0.16}$As$_2$ \cite{A114755} but distinct from the more three-dimensional case of BaFe$_{1.9}$Ni$_{0.1}$As$_2$ \cite{A121354}, the resonant spin correlations show no ${\bf Q}\cdot{\bf c}$ dependence beyond that associated with the product of the Fe magnetic form factor squared and the varying projection of the instrumental resolution volume along $\bf c$  (solid line). 

\begin{figure}
\includegraphics[width=60mm,angle=90]{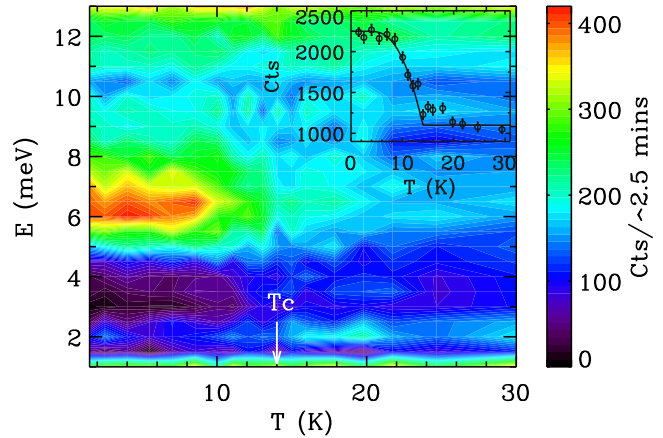}
\caption{(color) The energy scan at ${\bf Q}=(0.46,0.46,0.66)$ as a function of temperature to show the association between the spin resonance and superconductivity in FeSe$_{0.4}$Te$_{0.6}$. The sample turned background has been subtracted from the data. The inset shows the integrated intensity of the resonance between 5 and 8 meV as a function of temperature, and the line is a fit to the mean field theory with $T_c=14$ K.}
\label{fig4}
\end{figure}
Figure~\ref{fig4} shows the $\hbar\omega$-$T$ dependence of magnetic scattering at the nesting vector. The spin resonance and the associated spin gap appear along with superconductivity for $T<T_c=14$ K. There is no detectable softening of the resonance energy upon heating, indicating that it remains a characteristic energy scale in the normal state spectrum. In the inset, the temperature dependence of the integrated intensity of the resonance resembles an order 
parameter for the superconducting transition as in unconventional cuprate \cite{ybco_resn} and heavy fermion \cite{U123_resn,Co115_stock} superconductors.
The ratio between the resonance energy and the superconducting transition temperature $\hbar\Omega_0/k_BT_c=5.3$ for our FeSe$_{0.4}$Te$_{0.6}$ sample is 
larger than 2-4 reported for heavy fermion superconductors \cite{U123_resn,Co115_stock}, 4.3 for Ba$_{0.6}$K$_{0.4}$Fe$_2$As$_2$ \cite{A073932} and  4.5 for
BaFe$_{1.84}$Co$_{0.16}$As$_2$ \cite{A114755}, but comparable to the canonical value of 5 for cuprate superconductors \cite{bourges}. It also follows a general trend for a wide range of quantum condensation phenomena \cite{UemuraII}.

Turning now to the theoretical interpretation of the data, we note that the ${\bf Q}$ and $E$ dependent $\chi''({\bf Q}, E)$ measured through magnetic neutron scattering in the superconducting state reflects the symmetry of the superconducting gap function \cite{jmr_chi}. 
Due to the emblematic $s_{\pm}$ coherence factors for the interband processes, 
$1 + \frac{\Delta^2}{E_q^2} $, the creation of a pair of Bogoliubov-deGennes (BdG)
 quasiparticles is {\em enhanced}, in contrast 
to being suppressed as in a conventional $s$-wave state, where the 
corresponding coherence factor is $1 - \frac{\Delta^2}{E_q^2}$ \cite{B014790}. This leads to
a divergence in the imaginary part of $\chi_0({\bf Q}_n, E \rightarrow 2 \Delta)$ \cite{A041793}.
In addition, interactions pull the resonant 
peak below $2 \Delta$, creating a ``bound state''
of two BdG quasiparticles within the superconducting gap. 

We now demonstrate that this rather simple theoretical picture is consistent with the present data.
The explicit calculation employs an RPA-type scheme and uses a two-band model, 
with one hole and one electron parabolic 2D bands (inset to Fig.~\ref{fig1}(d)).
The band parameters are from ARPES measurements \cite{B011299}.
The position of the resonance peak 
is mainly controlled by the strength of the interaction while
the eccentricity determines its width in ${\bf Q}$ - extracting 
those parameters from the fits is a well-defined procedure. Here we use
$2\Delta_0 = 7.5$~meV ($2\Delta_0/k_BT_c \approx 6.1$);
higher than the BCS value but within the range measured in pnictides \cite{A070398}.
The eccentricity of the electron band 
is around $0.83$, the interaction strength is set to $0.3$ in units of inverse DOS, 
and areas of the hole and the electron pockets are roughly similar. 
More generally, reasonable fits are obtained for eccentricity and interaction in the
range 0.83-0.90 and 0.26-0.34, respectively.
Following the standard procedure, a small fixed $\chi''$ ($ \approx 0.1 \Delta_0$)
was added to smooth out the numerics. For comparison to Fig. 1c we subtracted the theoretical 30 K normal state intensity and convolved with the instrumental resolution. 
The result captures the essential physics of the resonance, 
as shown in Fig.~\ref{fig1}(d).

The main insight gained by this calculation is that the good
fit to the observed shape and position of the resonance necessarily calls for 
significant deviations from perfect nesting. In agreement with the
ARPES data, we find best fits for
a circular hole band and an elliptical electron band. This explains 
the absence of an antiferromagnetic spin-density-wave (SDW) 
ordering along the "nesting" vector in the 
normal state  \cite{A092058} - the peak in the spin susceptibility is highly sensitive to
deviations from perfect nesting \cite {A044678} and the 
combination of a 
depressed  peak and better-screened interactions 
can readily destabilize the SDW state. 
Overall, the above calculation
supports the $s_{\pm}$ superconducting state and the notion that 
superconductivity sets in after the itinerant SDW order has
been suppressed by deviations from perfect nesting.     

Since the discovery of the spin resonance in cuprate superconductors \cite{ybco_resn}, there has been much debate on whether it is associated with an intrinsic influence of superconductivity on spin correlations, or with a pre-existing collective mode of a nearby magnetic order enhanced by the loss of electron-hole pair damping in the superconducting state\cite{bourges}. According to the former scenario and the theory described above, the resonance should be a triplet - splitting linearly in an applied field. To determine the spin space multiplicity of the resonance, we carried out a constant-$\hbar\omega$ scan in a field of 7 Tesla. No splitting is directly visible in the data shown in Fig. 2b. Fitting these data to a triplet (doublet) places an upper limit of 1.3 meV (1.2 meV) on the overall level spacing. For comparison Zeeman splitting of a spin multiplet with $g=2$ would amount to $g\mu_B H=0.81$ meV. Higher fields may help to overcome the zero field broadening and determine the multiplicity of the resonance. 

In summary, we observe a strong quasi-two-dimensional spin resonance at the energy $\hbar\Omega_0=6.5$ meV and the wavevector ${\bf Q}_n=(\frac{1}{2}\frac{1}{2}L)$ in superconducting $\rm FeSe_{0.4}Te_{0.6}$. The peak has finite half widths in momentum and energy of $0.05(1)\AA^{-1}$ and $\hbar\Gamma=1.25(5)$~meV respectively. These experimental results are consistent with theoretical predictions for a $s_{\pm}$ superconducting pairing function \cite{A041793,B014790,A044678}. The normal state spin excitations appears to be itinerant antiferromagnons. Despite a different critical wavevector ${\bf Q}_m$ in the antiferromagnet parent compound $\rm Fe_{1+y}Te$, imperfect nesting of hole and electron Fermi surfaces separated by ${\bf Q}_n$ appears to lie behind $s_{\pm}$ superconductivity in $\rm FeSe_{0.4}Te_{0.6}$ as in the other Fe-based superconductors. 

Work at Tulane was supported by the NSF under grant DMR-0645305 for materials, the DOE under DE-FG02-07ER46358 for graduate students, and by the Research Corporation. Work at JHU is supported by the DOE under DE-FG02-08ER46544. SPINS is in part supported by NSF under agreement DMR-0454672.

Note: After completing our experiments at BT7 and SPINS on March 9, 2009, a related preprint describing neutron scattering experiments on a mixture of FeSe$_{0.45}$Te$_{0.55}$ and FeSe$_{0.65}$Te$_{0.35}$ came to our attention\cite{B042178}.


\begin{thebibliography}{39}
\expandafter\ifx\csname natexlab\endcsname\relax\def\natexlab#1{#1}\fi
\expandafter\ifx\csname bibnamefont\endcsname\relax
  \def\bibnamefont#1{#1}\fi
\expandafter\ifx\csname bibfnamefont\endcsname\relax
  \def\bibfnamefont#1{#1}\fi
\expandafter\ifx\csname citenamefont\endcsname\relax
  \def\citenamefont#1{#1}\fi
\expandafter\ifx\csname url\endcsname\relax
  \def\url#1{\texttt{#1}}\fi
\expandafter\ifx\csname urlprefix\endcsname\relax\def\urlprefix{URL }\fi
\providecommand{\bibinfo}[2]{#2}
\providecommand{\eprint}[2][]{\url{#2}}

\bibitem[{\citenamefont{Kamihara et~al.}(2008)\citenamefont{Kamihara, Watanabe,
  Hirano, and Hosono}}]{Kamihara2008}
\bibinfo{author}{\bibfnamefont{Y.}~\bibnamefont{Kamihara et al.}},
  \bibinfo{journal}{J.\ Am.\ Chem.\ Soc.} \textbf{\bibinfo{volume}{130}},
  \bibinfo{pages}{3296} (\bibinfo{year}{2008});
\bibinfo{author}{\bibfnamefont{X.~H.} \bibnamefont{Chen et al.}},
  \bibinfo{journal}{Nature} \textbf{\bibinfo{volume}{453}},
  \bibinfo{pages}{761} (\bibinfo{year}{2008}{\natexlab{a}});
\bibinfo{author}{\bibfnamefont{G.~F.} \bibnamefont{Chen et al.}},
  \bibinfo{journal}{Phys. Rev. Lett.} \textbf{\bibinfo{volume}{100}},
  \bibinfo{pages}{247002} (\bibinfo{year}{2008}{\natexlab{b}});
\bibinfo{author}{\bibfnamefont{Z.~A.} \bibnamefont{Ren et al.}},
  \bibinfo{journal}{Chinese Phys. Lett.} \textbf{\bibinfo{volume}{25}},
  \bibinfo{pages}{2215} (\bibinfo{year}{2008}).

\bibitem[{\citenamefont{Rotter et~al.}(2008)\citenamefont{Rotter, Tegel, and
  Johrendt}}]{A054630}
\bibinfo{author}{\bibfnamefont{M.}~\bibnamefont{Rotter et al.}},
  \bibinfo{journal}{Phys. Rev. Lett.} \textbf{\bibinfo{volume}{101}},
  \bibinfo{pages}{107006} (\bibinfo{year}{2008}).

\bibitem[{\citenamefont{Wang et~al.}(2008)\citenamefont{Wang, Liu, Lv, Gao,
  L.X.Yang, Yu, Li, and Jin}}]{A064688}
\bibinfo{author}{\bibfnamefont{X.}~\bibnamefont{Wang et al.}},
  \bibinfo{journal}{Solid State Comm.} \textbf{\bibinfo{volume}{148}},
  \bibinfo{pages}{538} (\bibinfo{year}{2008}).

\bibitem[{\citenamefont{Hsu et~al.}(2008)\citenamefont{Hsu, Luo, Yeh, Chen,
  Huang, Wu, Lee, Huang, Chu, Yan et~al.}}]{A072369}
\bibinfo{author}{\bibfnamefont{F.-C.} \bibnamefont{Hsu et al.}},
 \bibinfo{journal}{PNAS} \textbf{\bibinfo{volume}{105}},
  \bibinfo{pages}{14262} (\bibinfo{year}{2008}).

\bibitem[{\citenamefont{Singh}(2009)}]{B012149}
\bibinfo{author}{\bibfnamefont{D.~J.} \bibnamefont{Singh}},
  \bibinfo{journal}{arXiv:0901.2149}  (\bibinfo{year}{2009}).

\bibitem[{\citenamefont{Ma and Lu}(2008)}]{A033286}
\bibinfo{author}{\bibfnamefont{F.}~\bibnamefont{Ma}} \bibnamefont{and}
  \bibinfo{author}{\bibfnamefont{Z.~Y.} \bibnamefont{Lu}},
  \bibinfo{journal}{Phys. Rev. B} \textbf{\bibinfo{volume}{78}},
  \bibinfo{pages}{033111} (\bibinfo{year}{2008});
\bibinfo{author}{\bibfnamefont{J.}~\bibnamefont{Dong et al.}},
 \bibinfo{journal}{Europhys.\ Lett.}
  \textbf{\bibinfo{volume}{83}}, \bibinfo{pages}{27006} (\bibinfo{year}{2008}).

\bibitem[{\citenamefont{de~la Cruz et~al.}(2008)\citenamefont{de~la Cruz,
  Huang, Lynn, Li, Ratcliff, Zarestky, Mook, Chen, Luo, Wang et~al.}}]{A040795}
\bibinfo{author}{\bibfnamefont{C.}~\bibnamefont{de~la Cruz et al.}},
 \bibinfo{journal}{Nature}
  \textbf{\bibinfo{volume}{453}}, \bibinfo{pages}{899} (\bibinfo{year}{2008}).

\bibitem[{\citenamefont{Huang et~al.}(2008)\citenamefont{Huang, Qiu, Bao,
  Green, Lynn, Gasparovic, Wu, Wu, and Chen}}]{A062776}
\bibinfo{author}{\bibfnamefont{Q.}~\bibnamefont{Huang et al.}},
  \bibinfo{journal}{Phys. Rev. Lett.} \textbf{\bibinfo{volume}{101}},
  \bibinfo{pages}{257003} (\bibinfo{year}{2008}).

\bibitem[{\citenamefont{Goldman et~al.}(2008)\citenamefont{Goldman, Argyriou,
  Ouladdiaf, Chatterji, Kreyssig, Nandi, Ni, Budko, Canfield, and
  McQueeney}}]{A071525}
\bibinfo{author}{\bibfnamefont{A.}~\bibnamefont{Goldman et al.}},
  \bibinfo{journal}{Phys. Rev. B} \textbf{\bibinfo{volume}{78}},
  \bibinfo{pages}{100506(R)} (\bibinfo{year}{2008}).

\bibitem[{\citenamefont{Bao et~al.}(2008)\citenamefont{Bao, Qiu, Huang, Green,
  Zajdel, Fitzsimmons, Zhernenkov, Fang, Qian, Vehstedt et~al.}}]{A092058}
\bibinfo{author}{\bibfnamefont{W.}~\bibnamefont{Bao et al.}},
 \bibinfo{journal}{arXiv:0809.2058}
  (\bibinfo{year}{2008}).

\bibitem[{\citenamefont{Xia et~al.}(2009)\citenamefont{Xia, Qian, Wray, Hsieh,
  Chen, Luo, Wang, and Hasan}}]{B011299}
\bibinfo{author}{\bibfnamefont{Y.}~\bibnamefont{Xia et al.}},
  \bibinfo{journal}{arXiv:0901.1299}  (\bibinfo{year}{2009}).

\bibitem[{\citenamefont{Christianson et~al.}(2008)\citenamefont{Christianson,
  Goremychkin, Osborn, Rosenkranz, Lumsden, Malliakas, l.~S.~Todorov, Claus,
  Chung, Kanatzidis et~al.}}]{A073932}
\bibinfo{author}{\bibfnamefont{A.~D.} \bibnamefont{Christianson et al.}},
 \bibinfo{journal}{Nature}
  \textbf{\bibinfo{volume}{456}}, \bibinfo{pages}{930} (\bibinfo{year}{2008}).

\bibitem[{\citenamefont{Lumsden et~al.}(2008)\citenamefont{Lumsden,
  Christianson, Parshall, Stone, Nagler, MacDougall, Mook, Lokshin, Egami,
  Abernathy et~al.}}]{A114755}
\bibinfo{author}{\bibfnamefont{M.~D.} \bibnamefont{Lumsden et al.}},
 \bibinfo{journal}{Phys. Rev. Lett.}
  \textbf{\bibinfo{volume}{102}}, \bibinfo{pages}{107005}
  (\bibinfo{year}{2008}).

\bibitem[{\citenamefont{Chi et~al.}(2008)}]{A121354}
\bibinfo{author}{\bibfnamefont{S.}~\bibnamefont{Chi}} \bibnamefont{et~al.},
  \bibinfo{journal}{arXiv:0812.1354}  (\bibinfo{year}{2008});
\bibinfo{author}{\bibfnamefont{S.}~\bibnamefont{Li}} \bibnamefont{et~al.},
  \bibinfo{journal}{arXiv:0902.0813}  (\bibinfo{year}{2009}).

\bibitem[{\citenamefont{Uemura}(2009)}]{UemuraII}
\bibinfo{author}{\bibfnamefont{Y.~J.} \bibnamefont{Uemura et al.}},
  \bibinfo{journal}{Nature Mater.} \textbf{\bibinfo{volume}{8}},
  \bibinfo{pages}{253} (\bibinfo{year}{2009}).

\bibitem[{\citenamefont{Liu et~al.}(2009)}]{B040824}
\bibinfo{author}{\bibfnamefont{T.}~\bibnamefont{Liu}} \bibnamefont{et~al.},
  \bibinfo{journal}{arXiv:0904.0824}  (\bibinfo{year}{2009}).

\bibitem[{\citenamefont{Bao et~al.}(2000)\citenamefont{Bao, Pagliuso, Sarrao,
  Thompson, Fisk, Lynn, and Erwin}}]{bao00a}
\bibinfo{author}{\bibfnamefont{W.}~\bibnamefont{Bao et al.}},
  \bibinfo{journal}{Phys. Rev. B} \textbf{\bibinfo{volume}{62}},
  \bibinfo{pages}{R14621} (\bibinfo{year}{2000}), \bibinfo{note}{{\bf 67},
  099903(E) (2003)}.

\bibitem[{\citenamefont{Stock et~al.}(2008)\citenamefont{Stock, Broholm, Hudis,
  Kang, and Petrovic}}]{Co115_stock}
\bibinfo{author}{\bibfnamefont{C.}~\bibnamefont{Stock et al.}},
  \bibinfo{journal}{Phys. Rev. Lett.} \textbf{\bibinfo{volume}{100}},
  \bibinfo{pages}{087001} (\bibinfo{year}{2008}).

\bibitem[{\citenamefont{Bao et~al.}(1997)\citenamefont{Bao, Broholm, Aeppli,
  Dai, Honig, and Metcalf}}]{bao96c}
\bibinfo{author}{\bibfnamefont{W.}~\bibnamefont{Bao et al.}},
  \bibinfo{journal}{Phys. Rev. Lett.} \textbf{\bibinfo{volume}{78}},
  \bibinfo{pages}{507} (\bibinfo{year}{1997}).

\bibitem[{\citenamefont{Rossat-Mignod et~al.}(1991)}]{ybco_resn}
\bibinfo{author}{\bibfnamefont{J.}~\bibnamefont{Rossat-Mignod et al.}},
 \bibinfo{journal}{Physica C}
  \textbf{\bibinfo{volume}{185-189}}, \bibinfo{pages}{86}
  (\bibinfo{year}{1991});
\bibinfo{author}{\bibfnamefont{H.~A.} \bibnamefont{Mook}} \bibnamefont{et~al.},
  \bibinfo{journal}{Phys. Rev. Lett.} \textbf{\bibinfo{volume}{70}},
  \bibinfo{pages}{3490} (\bibinfo{year}{1993});
\bibinfo{author}{\bibfnamefont{H.~F.} \bibnamefont{Fong}} \bibnamefont{et~al.},
  \bibinfo{journal}{Nature} \textbf{\bibinfo{volume}{398}},
  \bibinfo{pages}{588} (\bibinfo{year}{1999});
\bibinfo{author}{\bibfnamefont{H.~F.} \bibnamefont{He}} \bibnamefont{et~al.},
  \bibinfo{journal}{Science} \textbf{\bibinfo{volume}{295}},
  \bibinfo{pages}{1045} (\bibinfo{year}{2002}).

\bibitem[{\citenamefont{Metoki et~al.}(1998)\citenamefont{Metoki, Haga, Koike,
  and Onuki}}]{U123_resn}
\bibinfo{author}{\bibfnamefont{N.}~\bibnamefont{Metoki et al.}},
  \bibinfo{journal}{Phys. Rev. Lett.} \textbf{\bibinfo{volume}{80}},
  \bibinfo{pages}{5417} (\bibinfo{year}{1998});
\bibinfo{author}{\bibfnamefont{O.}~\bibnamefont{Stockert et al.}},
  \bibinfo{journal}{Physica B} \textbf{\bibinfo{volume}{403}},
  \bibinfo{pages}{973} (\bibinfo{year}{2008}).

\bibitem[{\citenamefont{Bourges et~al.}(2005)}]{bourges}
\bibinfo{author}{\bibfnamefont{P.}~\bibnamefont{Bourges}} \bibnamefont{et~al.},
  \bibinfo{journal}{Physica C} \textbf{\bibinfo{volume}{424}},
  \bibinfo{pages}{45} (\bibinfo{year}{2005}).

\bibitem[{\citenamefont{Joynt and Rice}(1988)}]{jmr_chi}
\bibinfo{author}{\bibfnamefont{R.}~\bibnamefont{Joynt}} \bibnamefont{and}
  \bibinfo{author}{\bibfnamefont{T.~M.} \bibnamefont{Rice}},
  \bibinfo{journal}{Phys. Rev. B} \textbf{\bibinfo{volume}{38}},
  \bibinfo{pages}{2345} (\bibinfo{year}{1988}).

\bibitem[{\citenamefont{Mazin and Schmalian}(2009)}]{B014790}
\bibinfo{author}{\bibfnamefont{I.~I.} \bibnamefont{Mazin}} \bibnamefont{and}
  \bibinfo{author}{\bibfnamefont{J.}~\bibnamefont{Schmalian}},
  \bibinfo{journal}{arXiv:0901.4790}  (\bibinfo{year}{2009}).

\bibitem[{\citenamefont{Korshunov and Eremin}(2008)}]{A041793}
\bibinfo{author}{\bibfnamefont{M.~M.} \bibnamefont{Korshunov}}
  \bibnamefont{and} \bibinfo{author}{\bibfnamefont{I.}~\bibnamefont{Eremin}},
  \bibinfo{journal}{Phys. Rev. B} \textbf{\bibinfo{volume}{78}},
  \bibinfo{pages}{140509(R)} (\bibinfo{year}{2008});
\bibinfo{author}{\bibfnamefont{T.}~\bibnamefont{Maier}} \bibnamefont{and}
  \bibinfo{author}{\bibfnamefont{D.}~\bibnamefont{Scalapino}},
  \bibinfo{journal}{ibid.} \textbf{\bibinfo{volume}{78}},
  \bibinfo{pages}{020514(R)} (\bibinfo{year}{2008}).

\bibitem[{\citenamefont{Zhao et~al.}(2008)\citenamefont{Zhao, Liu, Zhang, Meng,
  Jia, Liu, Dong, Chen, Luo, Wang et~al.}}]{A070398}
\bibinfo{author}{\bibfnamefont{L.}~\bibnamefont{Zhao et al.}},
 \bibinfo{journal}{Chinese Phys. Lett.}
  \textbf{\bibinfo{volume}{25}}, \bibinfo{pages}{4402} (\bibinfo{year}{2008});
\bibinfo{author}{\bibfnamefont{H.}~\bibnamefont{Ding et al.}},
 \bibinfo{journal}{Europhys. Lett.}
  \textbf{\bibinfo{volume}{83}}, \bibinfo{pages}{47001} (\bibinfo{year}{2008}).

\bibitem[{\citenamefont{Cvetkovic and Tesanovic}(2008)}]{A044678}
\bibinfo{author}{\bibfnamefont{V.}~\bibnamefont{Cvetkovic}} \bibnamefont{and}
  \bibinfo{author}{\bibfnamefont{Z.}~\bibnamefont{Tesanovic}},
  \bibinfo{journal}{Europhys. Lett.} \textbf{\bibinfo{volume}{85}},
  \bibinfo{pages}{37002} (\bibinfo{year}{2008}).

\bibitem[{\citenamefont{Mook et~al.}(2009)\citenamefont{Mook, Lumsden,
  Christianson, Sales, Jin, McGuire, Sefat, Mandrus, Nagler, Egami
  et~al.}}]{B042178}
\bibinfo{author}{\bibfnamefont{H.~A.} \bibnamefont{Mook et al.}},
 \bibinfo{journal}{arXiv:0904.2178}
  (\bibinfo{year}{2009}).

\end{thebibliography}

\end{document}